\documentclass[a4paper]{jpconf}
\usepackage{graphicx}
\usepackage[cmex10]{amsmath}
\usepackage{mathtools}
\usepackage{mathrsfs}
\usepackage[cmex10]{amsmath}
\usepackage{bbm}
\usepackage{epsfig}
\usepackage{algorithm}
\usepackage{algorithmic}
\usepackage{color}
\definecolor{gray}{rgb}{0,0.1,0.5}
\usepackage{courier}
\usepackage{framed}
\usepackage{psfrag}

\usepackage[square,sort,comma,numbers, merge]{natbib}

\begin{document}
\title{A Study of Parallelizing O($N$) Green-Function-Based Monte Carlo Method
for Many Fermions Coupled with Classical Degrees of Freedom}

\author{Shixun Zhang}
\address{Department of Computer Science, University of Tsukuba, 1-1-1 Tennodai, Tsukuba, Ibaraki 305-8573,Japan}
\ead{sxzhang@cs.tsukuba.ac.jp}

\author{Shinichi Yamagiwa}
\address{Faculty of Engineering, Information and Systems
University of Tsukuba / JST PRESTO,
1-1-1 Tennodai, Tsukuba,
Ibaraki 305-8573, Japan}
\ead{yamagiwa@cs.tsukuba.ac.jp}

\author{Seiji Yunoki}
\address{Computational Condensed Matter Physics Laboratory, RIKEN ASI, Wako, Saitama, 351-0198 Japan, \\
Computational Materials Science Research Team, RIKEN AICS, 
Kobe, Hyogo, 650-0047 Japan, \\
CREST, Japan Science and Technology Agency, Kawaguchi, Saitama 332-0012, Japan}
\ead{yunoki@riken.jp}

\begin{abstract}
Models of fermions interacting with classical degrees of freedom are applied to a large variety of systems in condensed matter physics. 
For this class of models, Wei{\ss}e [Phys. Rev. Lett. {\bf 102}, 150604 (2009)] has recently proposed a very efficient numerical 
method, called O($N$) Green-Function-Based Monte Carlo (GFMC) method, where a kernel polynomial expansion technique is used 
to avoid the full numerical diagonalization of the fermion Hamiltonian matrix of size $N$, which usually costs O($N^3$) computational 
complexity.  
Motivated by this background, in this paper we apply the GFMC method to the double exchange model in three spatial dimensions. 
We mainly focus on the implementation of GFMC method using both MPI on a CPU-based cluster and Nvidia's Compute 
Unified Device Architecture (CUDA) programming techniques on a GPU-based (Graphics Processing Unit based) cluster. The time complexity of 
the algorithm and the parallel implementation details on the clusters are discussed. We also show the performance scaling for increasing Hamiltonian matrix size and increasing number of nodes, respectively.  The performance evaluation indicates that for a $32^3$ Hamiltonian a single GPU shows higher performance equivalent to more than 30 CPU cores parallelized using MPI.

\end{abstract}

\section{Introduction}\label{intro}
In solid state materials, we observe quite different physical properties such as superconductivity and magnetism. These different 
properties are explained by different behavior of electrons, which in turn is caused mostly by Coulomb interactions among electrons 
or/and interactions between electrons and other degrees of freedom such as phonon. 
Therefore, theoretical understanding of these interacting electrons is crucial to, e.g.,  
design new materials with rich functionalities. However, precisely due to this many-body nature of these interactions, it is usually 
very difficult to treat these systems analytically and even numerically. 

Among such interacting electron systems, models of electrons interacting with classical degrees of freedom can be applied to a large 
variety of systems. One of the most studied systems is the double exchange model for colossal magnetoregistive manganese 
oxides~\cite{yunoki, PhysRevB.58.6403,  PhysRevB.58.6414, yunoki2}. In the double exchange model, electrons can move around a lattice site under the influence of the localized 
classical spins via Hund's rule coupling, the Hamiltonian being given in Section~\ref{sec_formulization}. To solve this model, the Monte 
Carlo method is commonly used to carry out the importance sampling for the classical degrees of freedom after treating electron 
degrees of freedom by numerically diagonalizing the electron Hamiltonian of dimension $N$ (where $N$ is a system size)~\cite{yunoki, PhysRevB.58.6403,  PhysRevB.58.6414}. 
However, since at every Monte Carlo step one must fully diagonalize the Hamiltonian matrix to evaluate the partition function for a given 
configuration of the classical spins, it usually costs O($N^3$) numerical 
complexity. Because of this O($N^3$) scaling of the calculations, the accessible system size is strictly limited to up to hundreds to 
thousands sites, preventing us from studying important properties such as the Curie temperature of a ferromagnetic 
ordering~\cite{yunoki, PhysRevB.58.6403,  PhysRevB.58.6414}. 

In order to reduce the numerical complexity, Kernel Polynomial Method (KPM) is very advantageous~\cite{KPM}. Indeed, Motome and 
Furukawa have applied a Chebyshev expansion of the electron density of states for evaluating the partition function, which can reduce 
the numerical complexity down to O($N^2$)~\cite{motome}. They have also proposed that the numerical complexity can be further 
reduced to O($N$) by truncating the moment calculations for the Chebyshev expansion~\cite{furukawa}. Very recently, Wei{\ss}e has 
pointed out that the same numerical complexity of O($N$) can be achieved by using a Green-function-based Monte Carlo (GFMC) 
method, instead of directly evaluating the partition function for a given spin configuration~\cite{weisse_gfmc}. The GFMC method dramatically 
decreases the computational time by using only a few local Green's functions estimated by the Chebyshev expansion to calculate the 
change of the electron density of states. Moreover, this O($N$) GFMC method 
is more attractive than the one for directly evaluating the partition function with the same complexity of O($N$)~\cite{furukawa} 
because the GFMC method achieve the O($N$) complexity without any truncation~\cite{weisse_gfmc}. 

Motivated by these recent progress, here we apply the $O(N)$ GFMC method to the double exchange model. In this paper, we 
mainly focus on the 
implementation of the GFMC method using both MPI on a Central Processing Unit (CPU) cluster and Nvidia's Compute 
Unified Device Architecture (CUDA) programming techniques on a Graphics Processing Unit (GPU) cluster~\cite{CUDA}. The time complexity of 
the algorithm is estimated. The implementation details on a GPU cluster is also described since programming on a GPU cluster 
requires more 
attention on hardware aspects such as memory copy and communication between CPU and GPU. The performance evaluation 
indicates that the GPU program can archive over $10\times$ speedup as compared to the MPI parallelization on a CPU cluster.

This paper is organized as follows: Section~\ref{sec_formulization} explains the double exchange model and formulates the GFMC 
method, Section~\ref{sec_implmentation} describes the implementation details and the parallelization techniques, which are tested 
in Section~\ref{sec_validation}, Section~\ref{sec_performance} shows the performance on a single CPU processor as well as the whole cluster, 
and finally, Section~\ref{sec_conclusion} will give the conclusions.

\section{Model and Method Formulation }
\label{sec_formulization}
\subsection{Double Exchange Model}
The double exchange (DE) model describes electrons interacting the classical spins via Hund's rule coupling. The Hamiltonian 
is given by 
\begin{align}
H = - t \sum_{\langle i, j\rangle} \left( c_{i \sigma}^{\dag} c_{j \sigma} + {\rm H.c.} \right) + 
J_H \sum_{i, \alpha \beta} \vec {S}_{i} \cdot   c_{i \alpha}^{\dag} \vec \sigma_{\alpha\beta} c_{i \beta}  ,
\end{align}
where $c_{i \sigma}^{\dag}$ is a creation operator of electron at site $i$ and with spin $\sigma=(\uparrow,\downarrow)$, 
$\langle i, j\rangle$ runs over a pair of nearest neighbor sites $i$ and $j$, 
$\vec S_i$ is the classical spin at site $i$ with its normalization $|\vec S_i|=1$, and 
$J_H$ denotes the Hund's rule coupling with $J_H>0$. In the limit of infinite $J_H$ ($J_H\to\infty$), the Hamiltonian becomes
\begin{align}
H = -  \sum_{\langle i, j\rangle} t_{ij} \left( c_{i \sigma}^{\dag} c_{j \sigma} + {\rm H.c.} \right) 
\end{align}
with the hopping amplitude
\begin{align}
t_{ij} = \cos \frac{\theta_i - \theta_j}{2} \cos \frac{\phi_i - \phi_j}{2} + i \cos \frac{\theta_i + \theta_j}{2} \sin \frac{\phi_i - \phi_j}{2}, \label{ht}\\
\end{align}
where $\theta_i$ and $\phi_j$ denote the polar and the azimuthal angles of the classical spin $\vec S_i$, respectively.

The grand partition function of the DE model is written as
\begin{align}
Z = \prod_{i}^{N} \int d^3 \vec S_{i} {\rm Tr_e} \left[ {\rm exp} (- \beta H ( \{ \vec S_i \}) - \mu N  \right], 
\end{align}
where $\beta=1/T$ is inverse of temperature $T$, $\mu$ is the chemical potential, $N$ is the total number operator of electrons, 
and ${\rm Tr_e}[\cdots]$ indicates the trace over 
the electron degrees of freedom in the Fock space. 
The trace over classical spin degrees of freedom, i.e., $\prod_{i}^{N} \int d^3 \vec S_{i}P(  \{ \vec S_i \} )$, is evaluated by the Monte Carlo 
importance sampling with its weight $P(  \{ \vec S_i \} )$ for a given spin configuration $\{ \vec S_i\}$, 
\begin{align}
P(  \{ \vec S_i \} ) = {\rm Tr_e} \left[ {\rm exp} (- \beta H ( \{ \vec S_i \} - \mu N)  \right] = {\rm exp[-S_{eff} } (\{ \vec S_i \} )], 
\end{align}
where 
\begin{align}
    S_{\rm eff} ( \{ \vec S_i \} ) 
    & = - \sum_{i}^{N} {\rm log}(1 + e^{ - \beta [ \epsilon_i ( \{ \vec S_i \} ) - \mu]})  \\
    &= - \int {\rm log} (1 + e^{ - \beta [ E - \mu ] } ) \rho (E) dE,
\end{align}
$\epsilon_i$ is the $i$-th eigenvalue of Hamiltonian $H( \{ \vec S_i \} ) $, and $\rho (E)$ is the electron density of state (DOS) for a given spin 
configuration $\{ \vec S_i\}$. Using the Metropolis algorithm, the possibility 
$P ( \{\vec S_i\} \rightarrow \{\vec S_i^\prime\}) $ of accepting a new spin configuration $\{ {\vec S_i}^\prime \} $ is given by 
\begin{align}
P ( \{\vec S_i\} \rightarrow \{\vec S_i^\prime\}) 
= \frac{P (  \{\vec S_i^\prime\} )}{P ( \{\vec S_i\} )}  \nonumber
= e^{{\rm - S_{eff}} (  \{\vec S_i^\prime\} ) + {\rm S_{eff}} ( \{\vec S_i\} ) }.   \nonumber  
\end{align}
Therefore, the quantity $\Delta_{\rm seff}$ define by 
\begin{align}
\Delta_{\rm seff}= {\rm S_{\rm eff}} ( \{\vec S_i^\prime\} ) - {\rm S_{\rm eff}} ( \{\vec S_i\} ) =- \int {\rm log} (1 + e^{ - \beta [ E - \mu ] } ) ( \rho^\prime (E) - \rho (E) )dE \label{d_seff1}
\end{align}
is all we need to carry out the Monte Carlo calculation. Here, $ \rho^\prime (E) $ is  DOS for a given spin configuration $\{ \vec S_i^\prime\}$.
As mentioned in Section~\ref{intro}, we can exactly diagonalize $H ( \{ \vec S_i \})$ with O($N^3$) complexity to evaluate 
$\Delta_{\rm seff}$~\cite{yunoki, PhysRevB.58.6403,  PhysRevB.58.6414}, or we can use Kernel Polynomial Method (KPM) with O($N^2$) or O($N$) complexity to estimate 
the DOS directly~\cite{motome,furukawa}. However, the latter one with O($N$) complexity must truncate the moment calculations. 
Here, we use the recently proposed GFMC method~\cite{weisse_gfmc}, as will be explained below for completeness.  

\subsection{Green-function-based Monte Carlo (GFMC) method}
In this $O(N)$ GFMC method, we need to evaluate only several elements of the Green's function to calculate $\Delta_{\rm seff}$ 
given in Eq. (\ref{d_seff1}), provided that the change of spin configuration $\{ \vec S_i \}\to\{ \vec S_i^\prime \}$ is local, i.e, only a single 
spin at site $i$ is changed with the rest of spins unaltered.  We first give the definition of the Green's function $G(z)$ in the complex plane:  
\begin{align}
G (z) = \frac{1}{H - z} , \,\, z = E + i \epsilon ,
\end{align}
where $E$ is real and $\epsilon$ is a very small positive real. The Hamiltonian for a given spin configuration $\{ \vec S_i \}$ 
($\{ \vec S_i^\prime \}$) is denoted by $H$ ($H'$), and the Hamiltonian difference $\Delta$ is thus $\Delta = H^\prime - H$.  

The determinate of $G (z)  (H^\prime - z)$, i.e., 
\begin{align}
d(z) \coloneqq {\rm Det} \left[ G (z) (H^\prime - z) \right] &= {\rm Det } \left[ G(z) (H - z) + G(z) \Delta \right] 
= {\rm Det} \left[ \mathbbm{1} + G(z) \Delta \right] \label{green_delta} \\  
&=  {\rm Det} \left[ G (z)\right] \left[ ( H^\prime - z) \right] = \prod^{N}_{i} \frac{1}{ \epsilon_i - z } \prod^{N}_{i} { \epsilon^{\prime}_{i} - z } ,
\end{align}
has a special role in the GFMC method, 
where $\epsilon_i$ and $\epsilon^{\prime}_{i}$ are the $i$-th eigenvalues of $H$ and $H^\prime$, respectively. It is readily 
shown that
\begin{align}
\frac{1}{\pi} \lim\limits_{\epsilon \rightarrow 0}  {\rm Im} \frac{ {\rm d \,  log} (d(z)) }{dz} 
&= \frac{1}{\pi} \lim\limits_{\epsilon \rightarrow 0}  {\rm Im} \left( \sum_{i}^{N} \frac{1}{\epsilon_i - z} - \sum_{i}^{N} \frac{1}{\epsilon_i^{\prime} -z} 	\right) \nonumber \\
&= \sum_{i}^{N} \delta (E - \epsilon^\prime_i) - \sum_{i}^{N} \delta (E - \epsilon_i) \nonumber \\
&=  \rho (E)  - \rho^\prime (E). \label{dos_diff}
\end{align}
Thus, this equation in the left hand side can be used in Eq.~(\ref{d_seff1}), and $\Delta_{\rm seff}$ is now described using 
$d(z)$: 
\begin{align}
\Delta_{\rm seff} &=  \frac{1}{\pi} \lim\limits_{\epsilon \rightarrow 0} {\rm Im}  \int {\rm log} (1 + e^{ - \beta [ E - \mu ] } ) \frac{ {\rm d \,  log} (d(z)) }{dz} dE  \nonumber \\
&= \frac{\beta}{\pi}  \int \frac{1}{1 + e^{\beta (E - \mu ) }} \lim\limits_{\epsilon \rightarrow 0} {\rm Im \,} {\rm  log} (d(z))  dE. \label{d_seff2}
\end{align}

It is important to notice that we need only several elements of $G(z)$ to evaluate 
$d(z) = {\rm Det} \left[ \mathbbm{1} + G(z) \Delta \right]$.  As an example, here we consider the simple cubic lattice with the nearest neighbor 
hopping, and we assume that only one spin at site $o$ is changed: $\vec S_o\to\vec S_o^\prime$. In the cubic lattice, site $o$
has 6 nearest neighbors (NN) denoted by $\{ n, e, s, w, t, b\}$. Then,  the $\Delta$ matrix has a very simple form of 
{\small
\begin{align}
\Delta = 
\begin{bmatrix}
0 & 0 & 0 & \Delta_{n,o} & 0 & 0 & 0 \\
0 & 0 & 0 & \Delta_{e,o} & 0 & 0 & 0 \\
0 & 0 & 0 & \Delta_{s,o} & 0 & 0 & 0 \\
\Delta_{o,n} & \Delta_{o,e} & \Delta_{o,s} & 0 & \Delta_{o,w} & \Delta_{o,t} & \Delta_{o,b} \\
0 & 0 & 0 & \Delta_{w,o} & 0 & 0 & 0 \\
0 & 0 & 0 & \Delta_{t,o} & 0 & 0 & 0 \\
0 & 0 & 0 & \Delta_{b,o} & 0 & 0 & 0
\end{bmatrix}.
\end{align}
Therefore, to evaluate $d(z)$, only the following $7 \times 7$ Green's functions have to be calculated 
\begin{align}
G =
\begin{bmatrix}
G_{n,n} & G_{n,e} & G_{n,s} & G_{n,o} & G_{n,w} & G_{n,t} & G_{n,b} \\
G_{e,n} & G_{e,e} & G_{e,s} & G_{e,o} & G_{e,w} & G_{e,t} & G_{e,b} \\
G_{s,n} & G_{s,e} & G_{s,s} & G_{s,o} & G_{s,w} & G_{s,t} & G_{s,b} \\
G_{o,n} & G_{o,e} & G_{o,s} & G_{o,o} & G_{o,w} & G_{o,t} & G_{o,b} \\
G_{w,n} & G_{w,e} & G_{w,s} & G_{w,o} & G_{w,w} & G_{w,t} & G_{w,b} \\
G_{t,n} & G_{t,e} & G_{t,s} & G_{t,o} & G_{t,w} & G_{t,t} & G_{t,b} \\
G_{b,n} & G_{b,e} & G_{b,s} & G_{b,o} & G_{b,w} & G_{b,t} & G_{b,b}
\end{bmatrix}.
\end{align}
}

The computation can be further simplified by expanding $d(z)$ as the following: 

\begin{align}
d(z) &=\rm{det}(\mathbbm{1} + G(z) \Delta) \\
&=[ 1 + \sum_{j \in NN} \Delta_{jo}G_{oj}(z) ][1 + \sum_{j \in NN} \Delta_{oj}G_{jo}(z)] \\
&- G_{oo}[\sum_{j,k \in NN} \Delta_{jo}\Delta_{ok}G_{kj}(z)], 
\end{align}
where 
\begin{align}
\Delta_{jo} = \langle j | \Delta | o \rangle, G_{oj} = \langle o | G | j \rangle.
\end{align}
Moreover, using the following state $|v\rangle$:  
\begin{align}
| v \rangle =  \Delta | o \rangle = \sum_{j \in NN} \Delta_{jo} | j \rangle ,
\end{align}
$d(z)$ can be compactly expressed as
\begin{align}
d(z) = [1+G_{ov}(z)][1+G_{vo}(z)] - G_{oo}(z)G_{vv}(z). \label{d_z_green}
\end{align}
Notice that we now need only a $2 \times 2$ Green's function to evaluate $d(z)$.

Now, a question is how to calculate efficiently the local Green's functions $G(z)$. For this purpose, we use the KPM~\cite{KPM}, which 
can be efficiently implemented in a GPU cluster~\cite{zhang}. Using two types of Chebyshev polynomials ($m$: integer), 
\begin{align}
\left\{
\begin{aligned}
&T_m (x) = {\rm cos} \left[ m \, {\rm arccos} (x) \right] \\
&U_{m} (x) = \frac{ {\rm sin} \left[ (m + 1) {\rm arccos} (x) \right] }{ {\rm sin} \left[ {\rm arccos} (x) \right] } \label{cheby_series}
\end{aligned},
\right.
\end{align}
the diagonal elements of the Green's function are expanded as
\begin{align}
 G_{ii} (\tilde{w}+ i \epsilon) &= \frac{i}{\sqrt{1-\tilde{w}^2}} \left[ \tilde{\mu}_0 + 2 \sum_{m=1}^{M-1}  \tilde{\mu}_{m} {\rm T_m } (\tilde{w} ) \right]   
+2  \sum_{m=1}^{M-1}  \tilde{\mu}_{m} {\rm U_m } (\tilde{w} ) \nonumber \\
&=  \frac{i}{\sqrt{1-\tilde{w}^2}} \left[  \tilde \mu_0 + 2 \sum_{m=1}^{M-1} \tilde \mu_m {\rm exp}\left[-im {\rm arccos}(\tilde{w})\right]  \right] 
\label{green}.
\end{align}
Since the Chebyshev polynomials $T_m(x)$ and $U_m(x)$ requires that the argument $x $ should be within $[ -1, 1 ]$, we must 
renormalize the energy spectrum $E$ to $\tilde{w}$. The $\tilde{\mu}_m$ represents the $m$-th moment (defined below) after applying 
a kernel function, $\tilde{\mu}_m = \mu_m g_m$, where $g_m$ is the kernel function to eliminate Gibbs oscillations~\cite{KPM}. 
Here, we apply Lorenz kernel function which is defined as $g_m = {\rm sinh}[\lambda ( 1-m/M)] / {\rm sinh} (\lambda)$ with appropriate 
choice of $\lambda$~\cite{KPM}. The $m$-th moment $\mu_m$ is defined as 
\begin{align}
\mu_m & = \frac{1}{\pi} \lim\limits_{\epsilon \rightarrow 0}  {\rm Im}\int G_{ii} (\tilde{w} + i \epsilon) T_m ( \tilde{w}) d \tilde{w} \nonumber \\ 
& = \int \sum_n  \langle i | n \rangle \langle n | i \rangle \delta (\tilde{w} - \tilde{\epsilon}_n ) T_m ( \tilde{w}) d \tilde{w} \nonumber \\ 
& = \sum_n  \langle i | n \rangle \langle n | i \rangle   T_m (\tilde{\epsilon}_n) dE \nonumber \\ 
& =  \sum_n  \langle i | T_m (\tilde{H}) | n \rangle \langle n | i \rangle \nonumber \\ 
& = \langle i | T_m (\tilde{H}) | i \rangle,  
\end{align}
where $\tilde H$ is the renormalized Hamiltonian to fit the spectra within $[ -1, 1 ]$. 

Defining $| a_m \rangle = T_m (\tilde{H}) | i \rangle$ and thus $\mu_m = \langle i | a_m \rangle$,
the Chebyshev series in Eq.~(\ref{cheby_series}) yeilds the recursive relation of these vectors $|a_m\rangle$, namely, 
\begin{align}
\left\{
\begin{aligned}
&| a_0 \rangle = | i \rangle \\
&| a_1 \rangle = \tilde{H} | a_0 \rangle \\
&| a_{m} \rangle =2 \tilde{H} | a_{m-1} \rangle - | a_{m-2} \rangle \label{recursive}
\end{aligned}.
\right.
\end{align}
In addition, the number $M$ of coefficients is obtained with $M/2$ iterations if the following equation is used,
\begin{align}
&T_{2m-i} = 2 T_{m-i}   T_{m}  - T_{i} , \,\, i = 0,1 \\
&\mu_{2m-i} = 2 \langle r_{m-i} | r_{m} \rangle - \mu_i.  \label{mu2}
\end{align}

When the coefficients $\mu_m$ are all calculated, we can evaluate the Green's function using fast Fourier transform~\cite{KPM}.  
In Eq.~(\ref{green}), if we choose $$\tilde{w} = {\rm cos} \frac{\pi (k + \frac{1}{2})}{M}, \,\, (k=0, 1, ...,M-1)$$
the Green's function becomes 
\begin{align}
 G_{ii} (\tilde{w}+ i \epsilon) &= \frac{2i}{\sqrt{1-\tilde{w}^2}} \sum_{m=0}^{M-1} \mu_{m}^{\prime} {\rm exp}\left[ \frac{-i m \pi (k + \frac{1}{2})}{M} \right] , \label{green_short}
\end{align}
where 
\begin{align}
\mu^\prime_m = \left\{
\begin{aligned}
&\tilde{\mu}_0 / 2, \,\,m=0 \\
&\tilde{\mu}_m , \,\, m>0
\end{aligned}.
\right.
\end{align}
Let us now denote the summation part in Eq.~(\ref{green_short}) by $\chi_k$: 
$$\chi_k = \sum_{m=0}^{M-1} \mu_{m}^{\prime} {\rm exp}\left[ \frac{-i m \pi (k + \frac{1}{2})}{M}  \right].$$
It should be recalled that the following expression is required for the fast Fourier transformation~\cite{duhamel_fft}:
\begin{align}
\gamma_n = \sum_{m=0}^{M-1} c_{m} {\rm exp}\left[ \frac{-i 2 m \pi n}{M} \right], (m=0,1,\dots,M-1) 
\label{green_fft}
\end{align}
where \begin{align}c_{m} = \mu_{m}^{\prime}  {\rm exp}\left[ \frac{-i m \pi }{2M}  \right] \label{mu_to_mu_tilde} \end{align}
Using the following correspondence between $\chi_j$ and $\gamma_j$
$$
\left\{
\begin{aligned}
&\chi_{2j} = \gamma_j, \\
&\chi_{2j+1} = \gamma^{\ast}_{M-j-1}
\end{aligned}
\right. \,\,\,\, j=0,1,...,M/2-1 ,
$$
$\chi_{j}$ can be evaluated using the fast Fourier transformation, and thus the time complexity of calculating Eq.~(\ref{green_short}) 
reduces to O($M \, {\rm log} (M) $),  where $M$ is the number of moments kept. 

The off diagonal elements of the Green's function, $G_{ov}$ and $G_{vo}$, can be evaluated similarly if we use the follow mixed elements 
of the Green's function  
(note that $i$ below is the imaginary unit): 
\begin{align}
G_{o+v, o+v} &=(\langle v | + \langle o |) G ( | v \rangle + | o \rangle ) =G_{oo}+G_{vv}+G_{ov}+G_{vo} \\
G_{o+iv, o+iv} &=(- i \langle v | + \langle o |) G ( | o \rangle + i | v \rangle ) = G_{oo} +G_{vv}+ iG_{ov}-iG_{vo}, 
\end{align}
$G_{ov}$ and $G_{vo}$ are now expressed by the diagonal elements of the Green's functions: 
\begin{align}
&G_{ov} = \frac{1}{2} [ G_{o+v, o+v} - G_{oo} -G_{vv} + i (G_{oo}+G_{vv}-G_{o+iv, o+iv}) ], \\
&G_{vo} = \frac{1}{2} [ G_{o+v, o+v} - G_{oo} -G_{vv} + i (G_{o+iv, o+iv} - G_{oo} - G_{vv}) ].
\end{align}
Using these four elements of the $2\times2$ Green's function, we can readily calculate $d(z)$ and thus $\Delta_{\rm seff}$ can 
be evaluated.

\begin{figure}
\centering
\epsfig{file=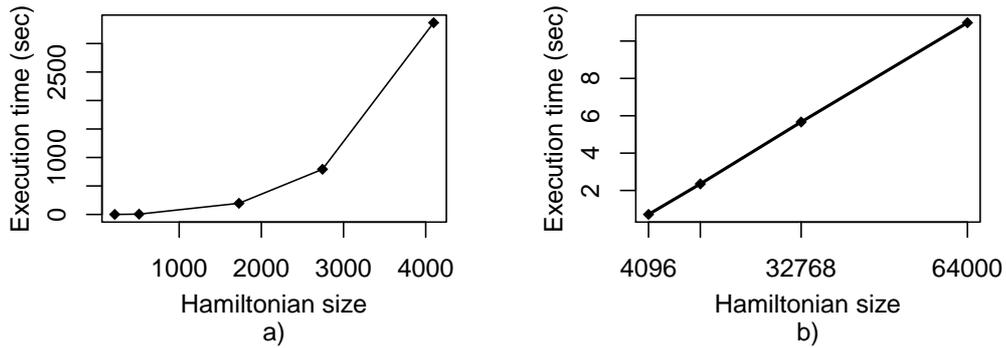, bb=20 20 450 150, width=12cm} 
\caption{
Hamiltonian size dependence of the execution time for (a) the exact diagonalization method~\cite{yunoki, PhysRevB.58.6403,  PhysRevB.58.6414} and (b)the GFMC method. 
Here, for simplicity, we performed only 10 Monte Carlo trial flips of spins, and $M$ is fixed for different Hamiltonian sizes. 
We can clearly see that the time consumption for the exact diagnolization method 
roughly follow the complexity of $O(N^3)$, but for the GFMC method it is almost linear.   }\label{fig_time_scaling} 
\end{figure}

Finally, it should be noted that if Hamiltonian matrix $H$ is stored in a compression format, the time complexity of 
Eq.~(\ref{recursive}) is $O(NM)$, where $N$ is the dimension of $H$.  As mentioned above, the time complexity to calculate 
Eq.~(\ref{green}) is $O(M \, {\rm log} (M))$, and thus the total time complexity is expected to be $O(NM) + O( M \, {\rm log} (M) )$. 
When $M$ is fixed, the time complexity should scale linearly with the dimension of $H$. Indeed, we find, as shown in 
Fig.~\ref{fig_time_scaling}, that the execution time is approximately proportional to the Hamiltonian size $N$.

\section{Implementation and Parallelization Schemes }
\label{sec_implmentation}

\subsection{Algorithms}
\label{sec_alg}
For a given temperature $T$, we use Algorithm~\ref{alg_gfmc} to calculate the magnetization $M$ for the classical spins through the average 
over $S$ Monte Carlo sweeps (the loop from line~\ref{alg_s_loop}), where each sweep corresponds to $N$ spin trail flips 
(the loop from line~\ref{alg_n_loop}). Since the direction of the magnetization is trivial, the magnetization $M$ is defined here 
as the length of the total spin vector.  

In the implementation, we apply the Metropolis method (line \ref{metropolis} in the Algorithm~\ref{alg_gfmc}) to determine whether a trail flip is accepted by comparing 
with a random number between 0 and 1. 

Section~\ref{sec_formulization} has illustrated how to calculate the $\Delta_{\rm seff}$ (from line \ref{alg_delta_matrix} to \ref{alg_d_seff}) 
using the GFMC method.  Especially, the calculation of the expansion coefficients $\mu_m$ plays a curial role (line \ref{alg_coefficients}). 
This part occupies most of the execution time since the recursion [denoted by Eq.~(\ref{recursive})] involves intensive matrix-vector 
multiplication (MVM) with complexity of $O(N)$. Note, however, that GPU is an ideal platform to parallelize MVM, because the 
multiplications between the rows and the vector could be distributed to hundreds of streaming processors. Therefore, we focus on a GPU 
implementation with the highly parallelism and expect a large speedup factor as compared with the CPU one. 
 
\begin{algorithm}
\caption{Calculate the magnetization as a function of temperature}          
\label{alg_gfmc}
\begin{algorithmic}[1]
    \REQUIRE Integer $S$ to represent number of Monte Carlo sampling sweeps
    \REQUIRE Hamiltonian matrix $H$ of dimension $N \times N$
    \REQUIRE Scalar $M$ to store accumulated magnetization
    \FOR{$ i = 1 \to S$}  \label{alg_s_loop}
    	    \FOR{$ j = 1 \to N$} \label{alg_n_loop}
    			\STATE Randomly choose site $o$ and change the spin randomly 
        		\STATE Calculate the modification matrix $\Delta \gets H - H^\prime$ \label{alg_delta_matrix} using Eq.~(\ref{ht})
       			\STATE Get vector $\overrightarrow{v} \gets \Delta \overrightarrow{o}$
       			\STATE Calculate the coefficients $\tilde{\mu}^{(o)}_m, \tilde{\mu}^{(v)}_m, \tilde{\mu}^{(o+iv)}_m, \tilde{\mu}^{(o-iv)}_m$ applied Lorentz kernel function $g_m$, where $\tilde{\mu}^{(V)}_m = \langle V | T_m (\tilde{H}) | V \rangle g_m$ \label{alg_coefficients}
       			\STATE Calculate 4 elements of the Green's function $G_{oo}, G_{vv}, G_{o+iv,o+iv}$ and $G_{o-iv,o-iv}$ using Eq.~(\ref{green})
       			\STATE Calculate $d(z)$ using Eq.~(\ref{d_z_green})
        		\STATE Calculate $\Delta_{\rm seff}$ using Eq.~(\ref{d_seff2}) \label{alg_d_seff}
           		\IF{ $e^{- \Delta_{\rm seff}} > $ rand() }  \label{metropolis}
           		 \STATE Accept the new spin configuration.
           		\STATE Update $H$ :  $H \gets H + \Delta$
           		\ENDIF
           \ENDFOR
           \STATE $M = M + \frac{\left| \sum^{N}_{i} \vec{S}_i \right|}{N} $ \label{alg_s_sum}
        \ENDFOR
        \STATE ${\rm Magnetization} \gets M / S $
\end{algorithmic}
\end{algorithm}

\subsection{Parallelization Frameworks}
\label{sec_parallel_framework}
The magnetization $M$ as a function of temperature $T$ is obtained through the average over a large number of Monte Carlo sweeps. 
If the system stays in the thermal equilibrium, the Monte Carlo sweeps are independent with each other. Therefore, we could divide the 
outside loop $S$ (line \ref{alg_s_loop}) into many threads. However, the warm up sweeps, which are necessary to achieve the thermal 
equilibrium, should be abandoned for taking the average, and this condition may prevent us from dividing $S$ to too many threads 
because some threads may have too few sweeps to reach thermal equilibrium.

In addition to this parallelism for the Monte Carlo sampling, the intensive matrix-vector and vector-vector multiplications in 
line \ref{alg_coefficients} can be effectively parallelized into multi-core CPU processors using OpenMP or many-core GPU processors 
using CUDA. In this paper, we focus on the implementation based on GPU to archive higher parallelism than multi-core CPU.

In our algorithm, we combine two parallelizations to achieve high efficiency, i.e., the Monte Carlo sweeps are divided into several 
MPI threads while in each thread we employ GPU to calculate matrix-vector operations.  

\subsection{Implementation on GPU}
\subsubsection{Compression Format of Hamiltonian Matrix}

Because the Hamiltonian matrix may be very large, e.g. $20^3 \times 20^3$, an effective compression technique must be introduced 
to reduce not only the memory consumption but also unnecessary memory access. The compression storage format of the Hamiltonian 
matrix may varies with different lattices and different physics models. For the simple cubic lattice used in this paper, each site has 6 
nearest neighbors that lead to 6 non-zero entries in each row. This case is very suitable for ELLPack format~\cite{ell} because no extra 
data need to be padded to the end of the rows.  ELLPack format needs two matrices to store the non-zero entries and 
column indices, respectively. On GPU device, the two matrices are stored in column major to trigger coalesced memory access.

\subsubsection{Arrangement of CUDA Kernel Functions}
 
One important issue that should be well addressed is the communication between GPU and CPU. Since the Hamiltonian matrix and the 
spin configurations may be updated after each iteration, if the Hamiltonian matrix are stored in CPU memory, the data transfer between 
CPU and GPU may significantly decrease the performance. This is especially true when the lattice size is small because the execution 
time for numerical calculation occupies a low percentage of the overall time consumption. According to our test, the simulation of 
$6\times6\times6$ cubic lattice on a 8x PCIe Tesla C2050 is about 30\% slower than on a 16x PCIe C2050. Therefore, the data transfer 
may not only decrease the performance, but also make the program to be more dependent on the bandwidth between the CPU and 
GPU. 

To avoid the memory transfer between CPU and GPU as much as possible, in our implementation the matrix $H$ and the spin 
configurations are kept in GPU memory. 

With these considerations, we propose a multi-kernel GPU implemenation shown in the Figure~\ref{fig_gpu_impl}, in which the kernel functions are indicated by "$<<<>>>$" and arranged as the following:

Function "FlipSpinAndCalculateDelta" flips a spin with a random direction and calculates the Hamiltonian matrix difference $\Delta$, which 
is used in the function "PrepareVector" to create four vectors $\{ | o \rangle, | v \rangle,  | o \rangle + i | v \rangle, | o \rangle - i | v \rangle \}$. 
For each vector, the function "CalculateCoefficient" calculates two expansion coefficients $\mu_m$ in every iteration. After all the 
coefficients are ready, function "PrepareForFFT" prepares the data for fast Fourier transformation (FFT), including applying Lorentz kernel 
function and transform Eq.~(\ref{green}) to meet the requirement of FFT. After performing FFT by function "PerformsFFT", we obtain 
$\Delta_{\rm seff}$ using function "CalculateDSEFF". If the flip is to be accepted, function "UpdateStatus" will update the spin configuration and the Hamiltonian matrix. 
After one MC sweep finishes, function "MeasureSpin" accumulates the spins to calculate the magnetization.

\begin{figure}
{
\begin{framed}
\texttt{  \scriptsize {
{
for $ i = 1 \to S$   \\  
\qquad for $ j = 1 \to N$   \\
\qquad  \qquad \textcolor{gray}  {  $//$ choose a spin randomly and flip a spin with a random direction } \\ 
\qquad  \qquad \textcolor{gray}  {  $//$  and calculate matrix $\Delta$} \\ 
\qquad \qquad FlipSpinAndCalculateDelta {$<<<>>>$} ();   \\ 
\qquad \qquad \textcolor{gray}  {  $//$ prepare vectors $| o \rangle, | v \rangle,  | o \rangle + i | v \rangle, | o \rangle - i | v \rangle$} \\ 
\qquad \qquad PrepareVector {$<<<>>>$} ();   \\ 
\qquad \qquad \textcolor{gray}  {  $//$ calculate the expansion coefficients for each Green's function} \\  
\qquad \qquad      for each vector in \{ $| o \rangle, | v \rangle,  | o \rangle + i | v \rangle, | o \rangle - i | v \rangle$ \} \\
\qquad \qquad   \qquad     for m=0 to M/2 \\
\qquad \qquad   \qquad   \qquad \textcolor{gray} {$//$ cacluate 2 coefficients for each iteration}
\qquad \qquad   \qquad   \qquad    CalculateCoefficient<<<>>> (); \\ 
\qquad \qquad  \qquad        end for \\
\qquad \qquad    end for \\
\qquad \qquad \textcolor{gray}  {  $//$ transform Eq.~(\ref{green}) to meet the requirement of FFT} \\ 
\qquad \qquad PrepareForFFT{$<<<>>>$} (); \\ 
\qquad \qquad \textcolor{gray}  {  $//$ perform FFT to obtain four Green's functions }\\ 
\qquad \qquad PerformsFFT{$<<<>>>$} (); \\ 
\qquad \qquad \textcolor{gray}  {  $//$ Calculate $\Delta_{\rm seff}$ }\\ 
\qquad \qquad CalculateDSEFF{$<<<>>>$} (); \\ 
\qquad \qquad \textcolor{gray}  {  $//$ If a trial spin flip is to be accepted, update the status }\\ 
\qquad \qquad UpdateStatus{$<<<>>>$} (); \\ 
\qquad \qquad \textcolor{gray}  {  $//$ Calculate the magnetization } \\
\qquad \qquad MeasureSpin{$<<<>>>$} (); \\
\qquad end for \\
end for 
}}}
\end{framed}}
\caption{Pseudocode of GPU implementation }\label{fig_gpu_impl} 
\end{figure}

\subsubsection{Implementation of the Recursion}
\begin{figure}
\centering
\epsfig{file=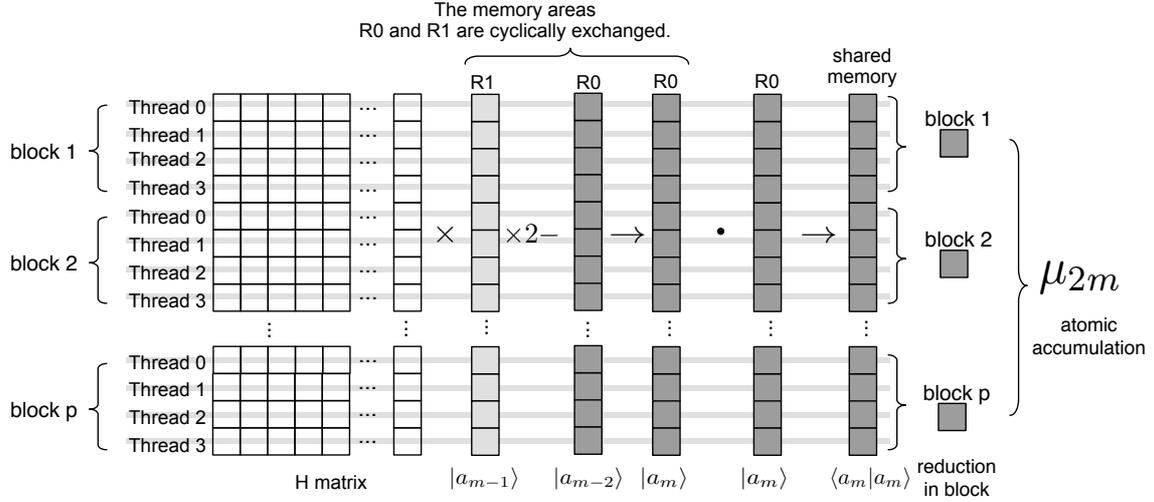, width=15cm} 
\caption{Implementation of function "CalculateCoefficient" to calculate the moments $\mu_{2m}$ }\label{fig_recursive} 
\end{figure}

The most important kernel function is "CalculateCoefficient", which involves the most intensive computation due to the recursive 
matrix-vector operations. Figure~\ref{fig_recursive}  demonstrates the implementation of this CUDA kernel function to calculate the 
coefficients $\mu_{2m}$. In CUDA architecture, the parallel running threads are divided into several thread groups, called "Blocks". 
For example, in Figure~\ref{fig_recursive} there are $p$ thread blocks and each block has 4 threads, indexed from 0 to 3. Each thread 
performs the multiplication between a row and the vector $| a_{m-1} \rangle$ that is stored in a 1D array R1. Since the produced vector 
$| a_m \rangle$ could be placed on R0 overwriting the vector $| a_{m-2} \rangle$, we only need two 1D memory arrays, R0 and R1, to 
perform calculations by exchanging their pointers after each iteration.

After the vector $| a_m \rangle $ is ready, a production $\langle a_m | a_m \rangle$ is performed, shared memory is applied to store the product result. $\langle a_m | a_m \rangle$ is a dot product, for CUDA, it is a reduction problem, which will be carried out in two steps, first, the product result in each block is accumulated to produce a scalar and then the scalar in each block is further accumulated to get coefficient $\mu_{2m}$ using Eq.~\ref{mu2}.  

This implementation could eliminate the most data transfers during the calculation. The bandwidth test shows the performance difference between a 8x and 16x PCIe based Tesla C2050 is very small (less than 6\%) while calculating a $6^3$ cubic lattice.  

\section{Validation}
\label{sec_validation}
\begin{figure}[t]
\centering
\epsfig{file=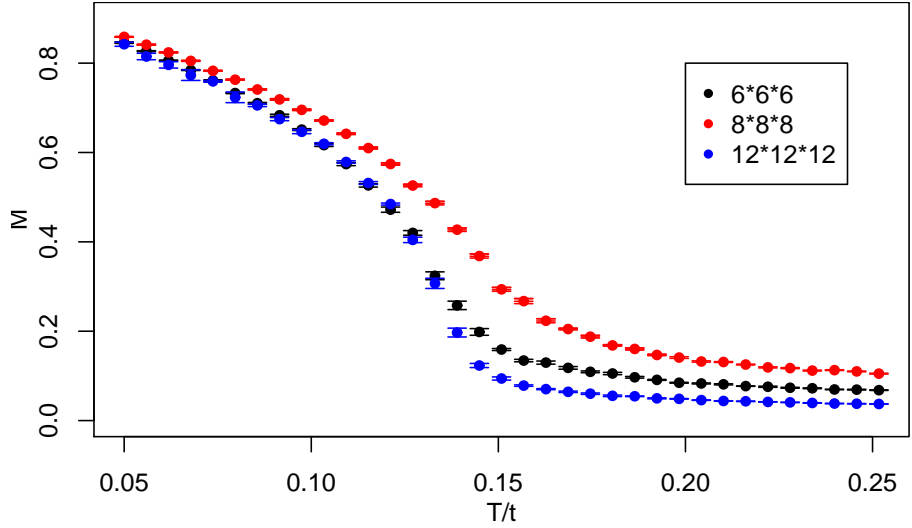, bb=20 30 450 280, clip=,  width=12cm} 
\caption{Magnetization $M$ as a function of temperature $T$ for different sizes (indicated in the figure) of the simple cubic lattice.}\label{fig_m_cubic} 
\end{figure}

In order to check the availability of our implementation, The DE model on the simple cubic lattice is simulated. The magnetization $M$ 
as a function of temperature $T$ is examined and the results are shown in Figure~\ref{fig_m_cubic}. The simulations are performed 
with the following parameters: Chebyshev truncation number is 256, chemical potential $\mu$ is 0, the average is taken over 5000 
MC sweeps.

\section{Performance evaluation}
\label{sec_performance}
\subsection{Experimental Setup}
The performance evaluation is performed on a cluster system that has 16 nodes of the NEC LX series connected with 40GByte/sec Infiniband network via the switch. Each node has two NVIDIA Tesla M2050 GPUs connected to the separate PCI Express buses and 12 CPU cores (Xeon E5645 2.4GHz). The OS of the node is CentOS~6 and the graphics driver with CUDA 4.0 is installed. The compiler for CPU is gcc-4.4.4. Two CPUs are tightly coupled with sharing the memory bus (totally 12 CPU cores in a node). Amount of total memory is 12GBytes per node. All experiments performed below use the compiler option "-O3" for CPU and GPU programs. A head node of the cluster is also connected to the cluster via the Infiniband network to serve the home file system in order to share the program resources. 

We evaluate the performance in two ways. The first is to measure the performance of a single CPU core and a single GPU with different Hamiltonian matrix sizes. The second is to evaluate the overall performance comparison between all CPU cores and all GPUs. 

\subsection{Performance Scaling on Multi-core CPU}
\begin{figure}
\centering
\epsfig{file=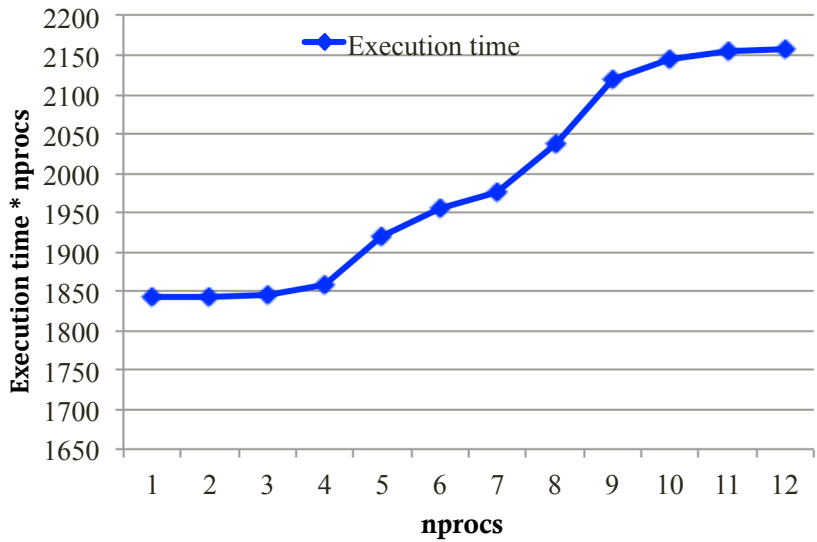, bb=0 0 450 280, clip=,  width=12cm} 
\caption{The performance scaling on multiple CPU cores within a node}\label{fig_perf_cpu_scaling} 
\end{figure}
Section~\ref{sec_alg} has explained that the process of calculating coefficients $\tilde{\mu}$ involves intensive MVM operations. The performance of MVM largely depends on the memory access speed. Therefore, the memory bandwidth plays an essential role in the overall performance. This conclusion is consistent with our mutli-core CPU performance scaling test shown in Figure~\ref{fig_perf_cpu_scaling}. 

This test is performed on a single node that has 2 CPUs, totally 12 cores and 12GB DDR2 800 memory with a peak bandwidth of 12.5GB/s. In order to evaluate the parallelization efficiency, we examined the quantity expressed by $y=execution\,\,time \times nprocs$, in which $nprocs$ represents the number of MPI threads. As the outside loop [from line 1] in Algorithm~\ref{alg_gfmc} is divided into mutliple MPI threads and there is very little communications among these threads, in an ideal case of parallelization, $y$ should be a constant. However, we have noticed that when the threads number becomes larger than 4, $y$ begins to increase fast, suggesting that the efficiency become to decrease rapidly. 

It indicates that the memory bandwidth becomes saturated when $nprocs$ grows to be larger than 4. Therefore, we can conclude that the bottleneck of this method is memory bandwidth. In this paper we resort to solve the problem using Tesla C2050 GPU with maximum bandwidth of 144GB/s~\citep{CUDA}.

\begin{figure}[t]
\centering
\epsfig{file=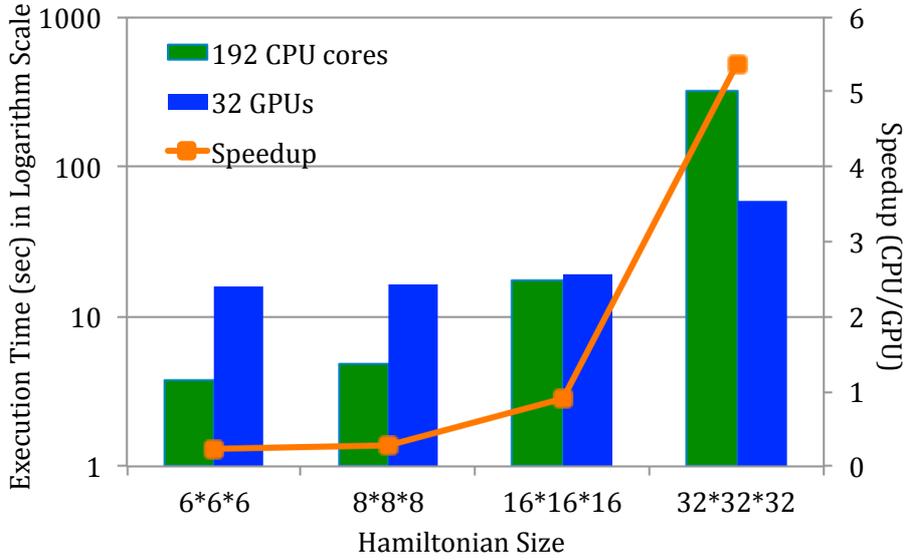, width=12cm} 
\caption{Performance scaling with Hamiltonian matrix size on cluster, the result is obtained with 38400 trail flips calculated and Chebyshev truncation number kept as 256}\label{fig_perf_hsize} 
\end{figure}

\subsection{Performance Scaling for Increasing Hamiltonian Size on Cluster}
Figure~\ref{fig_perf_hsize} shows the performance scaling on the cluster while the Hamiltonian size is increasing from $6^3$ to $32^3$. 38400 Monte Carlo trail flips are calculated while the number of Chebyshev moments are kept as 256. It can be observed that when the $H$ matrix is small, e.g. $6^3$ or $8^3$, 192 CPU cores are much faster than 32 GPUs. To explain the reason, let us divide the total time consumption of GPU program into two parts: a) the time for numerical calculation and b) the time for other works such as initializing GPU device, copying memory between CPU and GPU, and MPI communication, etc. Usually the latter part consumes very little time, but when the matrix size is small, comparing with the first part, time consumption of the latter part can not be neglected and significantly decrease the performance.  However, the speedup increases rapidly with increasing $H$ matrix size and in the best case, i.e. $32^3$, 32GPU is about over than 5 times faster than 192 CPU cores. If we define the speedup factor as the equivalent number of CPU cores comparing with one GPU. For Hamiltonian with size of $32^3$, this factor is about 30 (192/32 * 5). 

\begin{figure}[t]
\centering
\epsfig{file=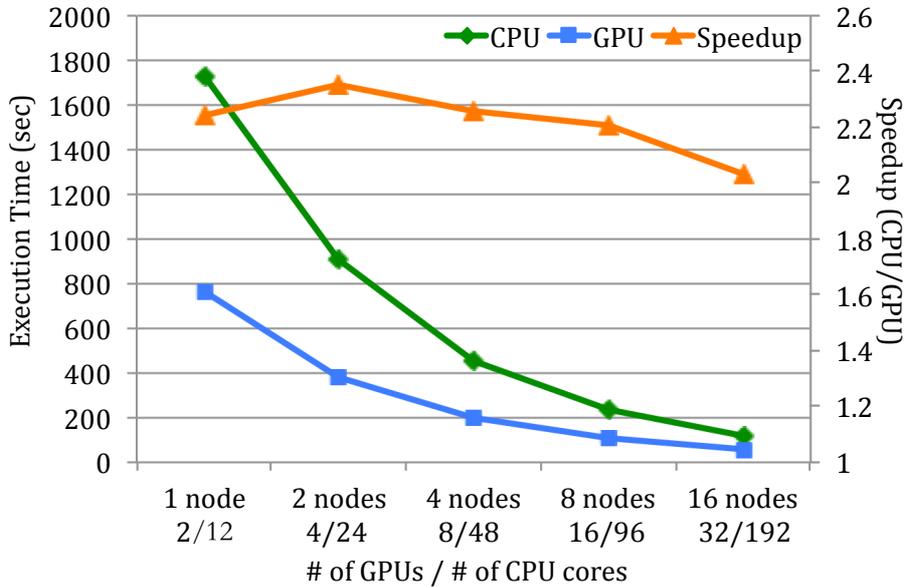, width=12cm}
\caption{Performance scaling for increasing number of nodes, each node has 2 Tesla C2050 and 12 Xeon 2.4GHz cores. 8000 Monte Carlo trail flips of a $20^3$ cubic lattice Hamiltonian are calculated with the Chebyshev truncation number kept as 256. The blue/green lines represent time consumption of CPU cores/GPUs}\label{perf_scaling_processor} 
\end{figure}

\subsection{Performance Scaling for Increasing Number of Nodes}
The performance scaling for increasing number of nodes is also very important reference for implementing the algorithm on a large cluster or even supercomputer. Here 8000 trail flips of a $20^3$ cubic lattice Hamiltonian is calculated on our cluster that has 2 Tesla C2050 and 12 Xeon 2.4GHz cores  in each node. The performance evaluation is shown in Figure~\ref{perf_scaling_processor}. As there is very little communication among MPI threads, it can be noticed that for both CPU and GPU implementation, the time consumption always decreases by half when the number of nodes is doubled , suggesting an ideal performance scaling with number of nodes. 

\subsection{Performance Considerations}
As shown in Figure~\ref{fig_perf_cpu_scaling}, the bottleneck of this algorithm is the memory bandwidth, GPU could solve the memory bandwidth limitation very well and shows high speedup ratio. Due to the nature that there is little communication among MPI threads, this algorithm is also capable to run on many compute nodes while keeping good parallelization efficiency. 

However, due to the thermal equilibrium problem  explained in Section \ref{sec_parallel_framework}, the of number of MPI threads is limited, in this case, a combination of parallelization techniques such as MPI, OpenMP and CUDA should be introduced. For example, the workload is firstly distributed into different nodes using MPI, and then in each node we use OpenMP to calculate the MVM, if the CUDA is capable, the OpenMP thread may employ GPU to calculate the MVM. The combination can  increase the parallelism and gain high performance. However, it should be pointed out that the process to combine these techniques may not be very straightforward since we have to take trade-offs in many occasions. 

\section{Conclusion and Future Work}
\label{sec_conclusion}
In this paper, we have provided the detailed formulation of the GFMC method for the DE model. 
We have proposed implementation methods to parallelize the GFMC method using MPI on CPU as well as CUDA on GPU. In order to 
eliminate the data transfer between CPU and GPU as much as possible, the program is implemented on GPU using several kernel 
functions. The performance evaluation indicates that the GPU implementation could effectively overcome the CPU memory bandwidth limitation and shows more than 30 speedup factor when Hamiltonian size is $32^3$. The performance scaling test for increasing number of processors indicates that the MPI parallelization is very effective for this algorithm. Finally we discussed more considerations on the performance for future optimizations.

For the future work, we will implement this method as a GPU based library that can be used for other lattice models.

\section*{Acknowledgement}
This work is partially supported by the Japan Science Technology Agency (JST) PRESTO program. And also this work is partially supported by KAKENHI (24300020) Grant-in-Aid for Scientific Research (B). We also would like to thank Dr. Alexander Weisse for insightful comments and suggestions to help implement the GFMC method.

\bibliographystyle{iopart-num}
\bibliography{references}

\end{document}